\documentclass{raa}            

\usepackage{graphicx,times}             
\usepackage{natbib}
\usepackage{amssymb,amsmath}
\bibpunct{(}{)}{;}{a}{}{,}

\usepackage[pagebackref=true]{hyperref}

\begin{document}

  \title{Flipping particles with nodal circulations by general relativity under the presence of an inner perturber}

   \volnopage{Vol.0 (202x) No.0, 000--000}      
   \setcounter{page}{1}          

   \author{G. C. de El\'{\i}a 
      \inst{1,2}
   \and M. Zanardi 
      \inst{1,2}
   \and C. F. Coronel 
      \inst{1,2}
    \and A. Dugaro
      \inst{1,2}
   }

   \institute{Instituto de Astrof\'{\i}sica de La Plata, CCT La Plata-CONICET-UNLP, Paseo del Bosque S/N (1900), La Plata, Argentina; {\it gdeelia@fcaglp.unlp.edu.ar}\\
        \and
            Facultad de Ciencias Astron\'omicas y Geof\'{\i}sicas, Universidad Nacional de La Plata, Paseo del Bosque S/N (1900), La Plata, Argentina\\
\vs\no
   {\small Received:; Accepted:}}

\abstract{
From a secular Hamiltonian up to the quadrupole level with general relativity (GR), we study nodal circulations with orbital flips of test particles of the Habitable Zone (HZ) around a solar-mass star, which are perturbed by an inner planetary-mass companion. Nodal circulations with orbital flips of an HZ test particle with eccentricity $e_2$ are possible for any mass $m_1$ and eccentricity $e_1$ of the inner perturber and a suitable inclination $i_2$. In particular, the greater the values of $m_1$ and $e_2$, the smaller the minimum extreme inclination $i_2$ capable of producing nodal circulations with orbital flips for each $e_1$. As long as nodal librations with orbital flips are not possible for any $i_2$, the greater the values of $m_1$, $e_1$, and $e_2$, the larger the region of the plane ($\Omega_2$, $i_2$) associated with nodal circulations with orbital flips. If nodal librations with orbital flips occur, the region of the plane ($\Omega_2$, $i_2$) referred to nodal circulations with orbital flips increases with a decrease in $m_1$ and $e_2$, and with an increase in $e_1$. We observe very good agreements between the analytical criteria and the N-body experiments for $m_1$ ranging from Earth-mass planets to super-Jupiters, and small and moderate $e_2$. The main discrepancies are found for high $e_2$, which are more evident with an increase in $m_1$ and $e_1$.
\keywords{planets and satellites: dynamical evolution and stability -- minor planets, asteroids: general -- relativistic processes -- methods: analytical -- methods: numerical}
}

   \authorrunning{G. C. de El\'{\i}a, M. Zanardi, C. F. Coronel, \& A. Dugaro}            
   \titlerunning{Flipping particles with nodal circulations by general relativity}

   \maketitle
%
\section{Introduction}           
\label{sect:intro}

For more than 100 years, it has been well known that the effects of general relativity (GR) produce an apsidal precession in the relative orbit of two close massive bodies \citep{Einstein1915}, which must be taken into account when analyzing the global dynamical properties of the system. Although \citet{Naoz2017} discussed the role of the GR on the dynamics of outer test particles perturbed by an inner Jupiter-mass planet around a solar-mass star, the study developed by \citet{Zanardi2018} represents the pioneer work that derives detailed analytical expresions capable of describing the dynamical properties of an outer test particle that orbits around two inner massive bodies including GR. Based on a secular Hamiltonian up to quadrupole level of the approximation, \citet{Zanardi2018} found an integral of motion of the system with GR, which is conserved over the evolutionary trajectory of each outer test particle. By analyzing the coupled evolution of the orbital inclination and the ascending node longitude, the authors showed that the GR leads to the existence of five different regimes of motion for an outer test particle. On the one hand, nodal librations, which can be associated with purely retrograde orbits and with orbital flips, where the inclination oscillates between prograde and retrograde values. On the other hand, nodal circulations, which can be experienced on purely prograde orbits, on purely retrograde orbits, as well as on flipping orbits. From this study, the orbital flips of an outer test particle in a restricted elliptical three-body problem with GR are associated with librations and circulations of the longitude of the ascending node, which differs from that reported in previous works that only consider classical Newtonian gravitation \citep{Ziglin1975, FaragoLaskar2010, Vinson2018, Lei2024}, which show that the orbital flips are exclusively correlated with nodal librations. Thus, a correct description of the dynamics of outer test particles in the framework of the elliptical restricted three-body problem must consider the GR effects.

Due to the requirement imposed by the transit technique, planets with high inclinations are difficult to detect. From this, theoretical works aimed at constraining the range of physical and orbital parameters of a system that lead to orbital flips of an outer body around a pair of massive bodies could help us better understand the true nature of a wide variety of planetary systems. Following this line of study, \citet{Coronel2024} recently used the analytical prescriptions derived by \citet{Zanardi2018} and \citet{Zanardi2023} and analyzed the role of GR in the nodal librations of test particles located in the habitable zone (HZ) around a solar-mass star and evolving under the influence of an inner-planetary-mass perturber. In that work, the authors combined analytical approaches and N-body experiments to analyze the dependence of orbital flips associated with nodal libration trajectories on the mass and eccentricity of the perturber, as well as on the eccentricity of the outer test particle.

In the present work, we are interested in continuing the research initiated by \citet{Coronel2024} on the dynamics of HZ test particles around a solar-mass star perturbed by an inner-planetary-mass companion with GR effects, focusing on orbital flips with nodal circulation trajectories. Our results, combined with those derived by \citet{Coronel2024}, will allow us to determine the full space of physical and orbital parameters that lead to oscillations of the orbital plane of HZ test particles from prograde to retrograde inclinations and back again due to GR. This line of investigation will help us better understand the results of future observational studies aimed at detecting planets in systems analogous to those proposed in our study scenarios.

Our research is organized as follows. In Sect.~\ref{sec:analytical_model}, we present the analytical model used to develop our study. In Sect.~\ref{sec:results}, we work on the basis of such analytical prescriptions and analyze the nodal circulation trajectories with orbital flips for different values of the mass and eccentricity of the inner perturber and of the eccentricity of the HZ test particle. Then, results derived from N-body experiments are shown with the aim of analyzing the robustness of our analytical model. Finally, we describe the discussions and conclusions of our research in Sect.~\ref{sec:conclusions}.

\begin{figure*}
   \centering
  \includegraphics[width=1
   \textwidth]{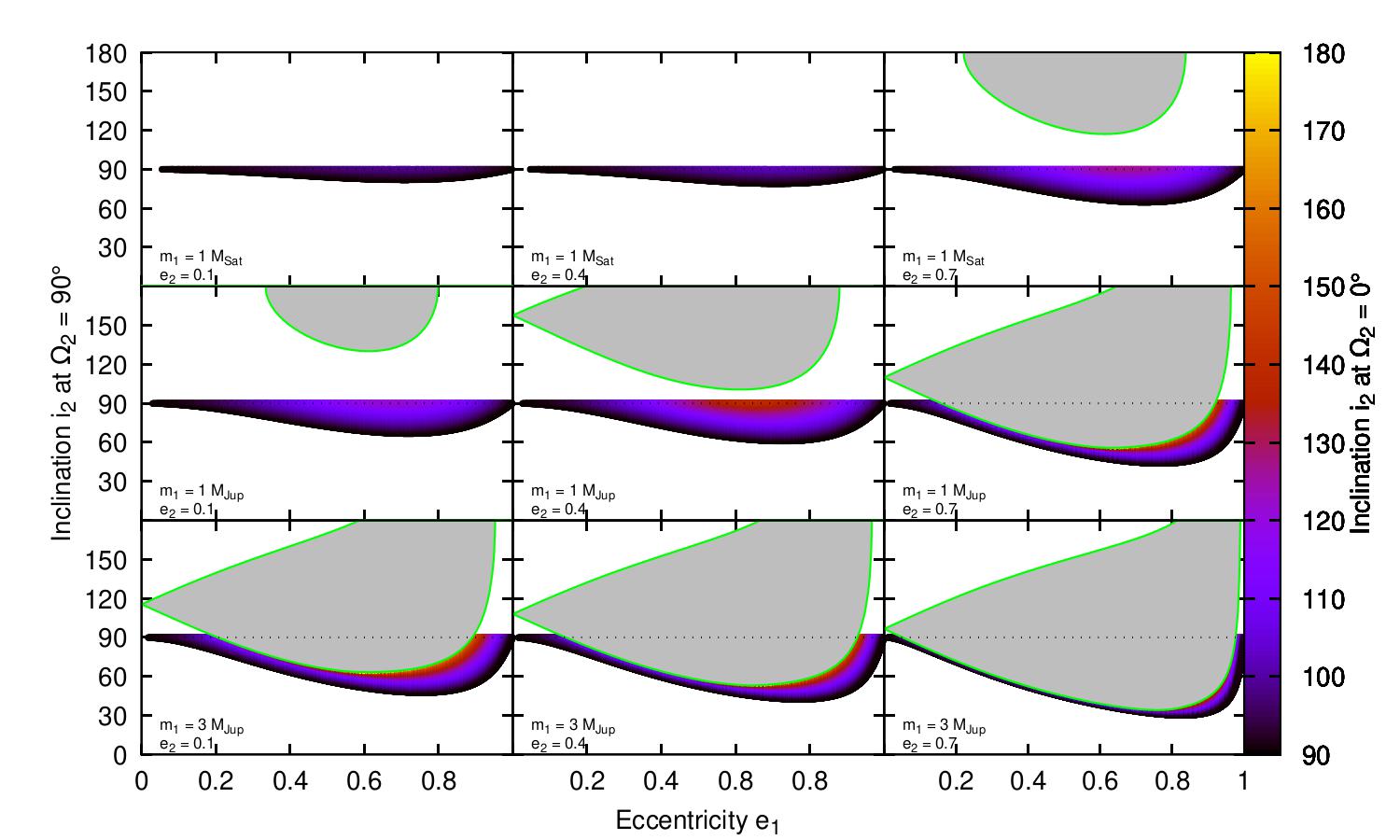}
   \caption{Regimes of motion of an HZ test particle in the plane $e_1$ vs. $i_2$ at $\Omega_2$ = $\pm$ 90$^{\circ}$ for different $m_1$ and $e_2$. In each panel, the colored region is associated with nodal circulations with orbital flips, where the color code refers to the retrograde value of $i_2$ at $\Omega_2$ = 0$^{\circ}$ reached by an evolutionary trajectory from a prograde $i_2$ at $\Omega_2$ = $\pm$ 90$^{\circ}$. Moreover, the nodal libration region is shown in gray, while the green curve represents the extreme inclinations $i^{\textrm{e}}_2$ obtained from Eq.~\ref{eq:cuadratica}. Finally, the nodal circulation region associated with purely prograde orbits ($i_2$ at $\Omega_2$ = $\pm$ 90$^{\circ}$ $<$ 90$^{\circ}$) and purely retrograde orbits ($i_2$ at $\Omega$ = $\pm$ 90$^{\circ}$ $>$ 90$^{\circ}$) is illustrated in white.   
 }
\label{fig:multiplot_grafico1}
\end{figure*}

\section{Analytical model}
\label{sec:analytical_model}

Here, we present the model used to analyze the dynamical behavior of an outer test particle in the elliptical restricted three-body problem with GR. In particular, we describe the prescriptions derived by \citet{Zanardi2018}, who developed an analytical approach adding GR effects to the secular Hamiltonian up to the quadrupole level of the approximation obtained by \citet{Naoz2017}. 

In agreement with the pioneering work carried out by \citet{Ziglin1975}, \citet{Naoz2017} showed that the Hamiltonian of an outer test particle up to the quadrupole level of the secular approximation in the elliptical restricted three-body problem is expressed by

\begin{eqnarray}
f_{\text{quad}} = \frac{\left(2 + 3e^2_{1}\right)\left(3\cos^2{i_{2}} - 1\right) + 15e^2_{1}\left(1 - \cos^2{i_{2}}\right)\cos{2\Omega_{2}}}{\left(1 - e^2_{2}\right)^{3/2}},
\label{eq:fquad}
\end{eqnarray}

\noindent{where} $e_1$ represents the eccentricity of the inner perturber, and $e_2$, $i_2$, and $\Omega_2$ refer to the eccentricity, the inclination with respect to the inner orbit, and the ascending node longitude of the outer test particle, respectively. It is important to mention that $\Omega_2$ is measured relative to the pericentre of the orbit associated with the inner perturber.  

By assuming classical Newtonian gravitation, Eq.~\ref{eq:fquad} allows to verify that an outer test particle can experience three different motion regimes based on the coupled evolution of $i_2$ and $\Omega_2$. On the one hand, a nodal circulation regime, where $\Omega_2$ is not constrained between two specific values and the evolutionary trajectories keep purely prograde or purely retrograde values of $i_2$, depending on its initial condition. On the other hand, a nodal libration regime, in which $\Omega_2$ is bounded between two values and $i_2$ oscillates between prograde and retrograde values and back again, defining an orbital flip. In particular, it is important to mention that the minimum and maximum inclinations that delimit the nodal libration region depend only on $e_1$ and are symmetric with respect to 90$^{\circ}$.

Later, \citet{Zanardi2018} analyzed the inclusion of GR effects in the elliptical restricted three-body problem for an outer test particle. GR produces apsidal precession of the orbit of the inner perturber \citep{Einstein1915}, for which the Hamiltonian model is formulated in a rotating coordinate system. From this, \citet{Zanardi2018} showed the existence of an integral of motion $f$ for an outer test particle in the elliptical restricted three-body problem with GR, which adopts the expression
 
\begin{eqnarray}
    f = f_{\text{quad}} + f_{\text{GR}},
    \label{eq:f}
\end{eqnarray}

\noindent{where} $f_{\text{quad}}$ is given by Eq.~\ref{eq:fquad} and $f_{\text{GR}}$ is given by

\begin{eqnarray}
    f_{\text{GR}} = \frac{48 k^2 \cos{i_{2}} \left(m_{1} + m_{\star}\right)^3 a_{2}^{7/2}\left(1 - e_{2}^2\right)^{1/2}}{m_{1}m_{\star}a_{1}^{9/2}c^2\left(1 - e_{1}^2\right)},
    \label{eq:fgr}
\end{eqnarray}

\noindent{where} $k^2$ is the gravitational constant, $c$ the speed of light, $m_{\star}$ and $m_1$ the mass of the star and the inner perturber, respectively, and $a_1$ and $a_2$ the semimajor axis of the inner perturber and the outer test particle, respectively. It is important to remark again that $\Omega_2$ in Eq.~\ref{eq:fquad} is relative to the pericenter of the inner perturber's orbit, so it is measured in a rotating system when GR is considered.

From this and based on the coupled evolution of $i_2$ and $\Omega_2$, \citet{Zanardi2018} found five different regimes of motion of an outer test particle, which are characterized by nodal librations, which can be associated with orbital flips or purely retrograde orbits, and by nodal circulations, which can be correlated with purely prograde orbits, purely retrograde orbits, or orbital flips.

For the nodal libration regime, \citet{Zanardi2018} showed that the extreme values of $\Omega_2$ are obtained for an inclination $i^{*}_{2}$ given by

\begin{eqnarray}
i_{2}^*=\arccos{\left(\frac{a_{2}^{7/2}\left(1 - e_{2}^2\right)^2A}{a_{1}^{9/2}\left(1 - e_{1}^2\right)\left(2 + 3e_{1}^2 - 5e_{1}^2\cos{2\Omega_{2}}\right)}\right)},
    \label{iestrella}
\end{eqnarray}

\noindent{where} $A$ adopts the expression 

\begin{eqnarray}
    A = - \frac{8k^{2}\left( m_{1} + m_{\star} \right)^{3}}{c^{2}m_{1}m_{\star}}.
    \label{eq:A}
\end{eqnarray}

\noindent{From} this and the use of the integral of motion $f$ given by Eq.~\ref{eq:f}, \citet{Zanardi2018} and \citet{Zanardi2023} showed that the extreme inclinations $i_{2}^{\textrm{e}}$ that lead to nodal librations of the outer test particle are obtained from  

\begin{eqnarray}
    \alpha\cos^2{i_{2}^{\textrm{e}}} + \beta\cos{i_{2}^{\textrm{e}}} + \gamma = 0,
    \label{eq:cuadratica}
\end{eqnarray}

\noindent{where} $\alpha$ and $\beta$ are always given by

\begin{eqnarray}
    \alpha = 1 + 4e^{2}_{1},
    \label{alfa}
\end{eqnarray}
\begin{eqnarray}
    \beta = - \frac{A\left(1 - e^2_{2}\right)^{2}a^{7/2}_{2}}{\left(1 - e^2_{1}\right)a^{9/2}_{1}}.
    \label{beta}
\end{eqnarray}

\noindent{If} Eq.~\ref{iestrella} has solution at $\Omega = 0^{\circ}$, $\gamma$ is calculated by

\begin{eqnarray}
    \gamma = \frac{\beta^2}{4\left(1 - e^2_{1}\right)} - 5e^{2}_{1}.
    \label{gamma_viejo}
\end{eqnarray}

\noindent{On} the contrary, $\gamma$ is given by the following expression 

\begin{eqnarray}
    \gamma = \beta - \alpha,
    \label{gamma_nuevo}
\end{eqnarray}

\noindent{from} which, the maximum extreme inclination $i_{2,\textrm{max}}^{\textrm{e}}$ is always equal to 180$^{\circ}$ and the minimum extreme inclination adopts a simple form given by

\begin{eqnarray}
i_{2,\textrm{min}}^{\textrm{e}} = \textrm{arccos} \left(1 - \frac{\beta}{\alpha} \right).
\label{eq:i2min_nueva_cuadratica}
\end{eqnarray}
According to this analysis, GR effects break the symmetry of the dynamical model, so the minimum and maximum inclinations that delimit the nodal libration region are no longer symmetrical with respect to 90$^{\circ}$, strongly depending on the physical and orbital parameters of the system.

In the present research, we are interested at analyzing nodal circulation trajectories for the outer test particle. In particular, our study is focused on nodal circulations with orbital flips, where the inclination $i_2$ oscillates between prograde and retrograde values. To do this, let $i_{2,0}$ be the inclination of the outer test particle at $\Omega_2 = 0^{\circ}$, and $i_{2,\text{min}}$ be the minimum inclination reached in its evolutionary trajectory, which is associated with $\Omega_2 = \pm 90^{\circ}$. Furthermore, let $i^{\text{r}}_{2,0}$ be a retrograde value of $i_{2,0}$. From this, a nodal circulation trajectory with orbital flip for the outer test particle is possible if there exists a value $i^{\text{r}}_{2,0}$ that leads to a prograde value of $i_{2,\text{min}}$. If this happens, and given that the integral of motion $f$ is conserved on the evolutionary trajectory of an outer test particle, $f(\Omega_2 = 0^{\circ}, i_2 = i^{\text{r}}_{2,0}) = f(\Omega_2 = \pm 90^{\circ}, i_2 = i^{\text{p}}_{2,\text{min}})$ is proposed, where $i^{\text{p}}_{2,\text{min}}$ represents a prograde value of $i_{2,\text{min}}$. From this, given $i^{\text{r}}_{2,0}$, $i^{\text{p}}_{2,\text{min}}$ is a solution of the following quadratic equation  

\begin{eqnarray}
    \alpha\cos^2{i^{\text{p}}_{2,\text{min}}} + \beta\cos{i^{\text{p}}_{2,\text{min}}} + \gamma^{\ddagger} = 0,
    \label{eq:cuadratica_flipping}
\end{eqnarray}

\noindent{where} $\alpha$ and $\beta$ are always given by Eqs.~\ref{alfa} and \ref{beta}, and $\gamma^{\ddagger}$ adopts the expression

\begin{eqnarray}
    \gamma^{\ddagger} = (e^{2}_{1}-1) \cos^2 i^{\text{r}}_{2,0} - \beta \cos^2 i^{\text{r}}_{2,0} - 5e^{2}_{1},
    \label{gamma_flipping}
\end{eqnarray}

By analyzing the coefficients of the Eq.~\ref{eq:cuadratica_flipping}, we show that the nodal circulation region with orbital flips of an outer test particle in the elliptical restricted three-body problem strongly depends on the physical and orbital parameters of the system bodies.   

\section{Results}
\label{sec:results}

Here, we study the trajectories of nodal circulation with orbital flips of an outer test particle in the elliptical restricted three-body problem with GR. In particular, we consider that all our systems of study are composed of a solar-mass star, an inner perturber with a semimajor axis $a_1 = 0.1$ au, and an outer test particle located at the HZ with a semimajor axis $a_2 = 1$ au. From this, we firstly use the analytical approach developed in Sect.~\ref{sec:analytical_model} in order to analyze the nodal circulation region with orbital flips of an HZ test particle for different values of $m_1$, $e_1$, and $e_2$. Finally, we carry out N-body experiments with the aim of studying the robustness of the results derived from the analytical theory. 

\begin{figure*}
   \centering
  \includegraphics[width=1
   \textwidth]{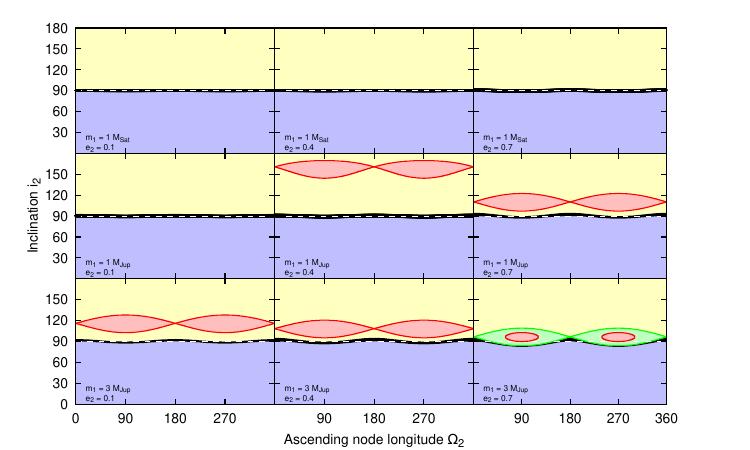}
   \caption{Regimes of motion of an HZ test particle in the plane $\Omega_2$ vs. $i_2$ for $e_1$ = 0.1 and different values of $m_1$ and $e_2$. In each panel, the region of nodal circulation with orbital flips is shown in black. Moreover, the blue and yellow regions represent the pairs ($\Omega_2$, $i_2$) that produce nodal circulation trajectories with prograde and retrograde orbits, respectively. The nodal libration regions associated with orbital flips and purely retrograde orbits are illustrated in green and red, respectively. Finally, the green curve transitions from nodal librations with orbital flips to nodal circulations, while the red curve represents an extreme case associated with nodal libration trajectories with purely retrograde inclinations.   
}
\label{fig:multiplot2_1}
\end{figure*}

\begin{figure*}
   \centering
  \includegraphics[width=1
   \textwidth]{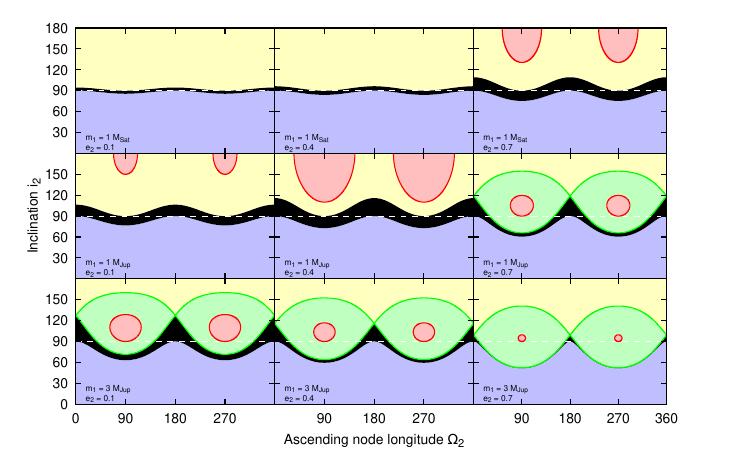}
   \caption{
Regimes of motion of an HZ test particle in the plane $\Omega_2$ vs. $i_2$ for $e_1$ = 0.4 and different values of $m_1$ and $e_2$. The references associated with the different regions and curves illustrated in each panel are described in
the caption of Fig.~\ref{fig:multiplot2_1}.   
}
\label{fig:multiplot2_2}
\end{figure*}

\subsection{Analytical Treatment}
\label{subsec:analytical_treatment}

From the solution of Eq.~\ref{eq:cuadratica_flipping}, our results show that an HZ test particle with eccentricity $e_2$ can experience nodal circulations with orbital flips for any eccentricity $e_1$ and mass $m_1$ of the inner perturber and an appropriate inclination $i_2$. Although this result is valid from sub-Earth-mass planets to super-Jupiters, it is worth remarking that an inner perturber with $m_1 \lesssim$ 1 M$_{\text{Nep}}$ only allows an HZ test particle to experience nodal circulations with orbital flips for $i^{\text{p}}_{2,\text{min}}$ and $i^{\text{r}}_{2,0}$ close to 90$^{\circ}$ for any value of $e_1$ and $e_2$. From such values of $i^{\text{p}}_{2,\text{min}}$ and $i^{\text{r}}_{2,0}$, the evolutionary trajectories have $i_2$ always close to 90$^{\circ}$ for any $\Omega_2$.

The color points shown in Fig.~\ref{fig:multiplot_grafico1} illustrate the $i^{\text{p}}_{2,\text{min}}$ values that lead to orbital flips with nodal circulations as a function of $e_1$ for different values of $m_1$ and $e_2$. The color code represents the value of $i^{\text{r}}_{2,0}$ reached by a nodal circulation trajectory from a given $i^{\text{p}}_{2,\text{min}}$. From top to bottom, the rows of this figure are associated with a perturber of 1 M$_{\text{Sat}}$, 1 M$_{\text{Jup}}$, and 3 M$_{\text{Jup}}$, while the columns correspond to $e_2$ values of 0.1, 0.4, and 0.7 from left to right. Moreover, the nodal libration region is shown in gray in each panel, while the green curves represent the extreme inclinations $i^{\text{e}}_2$ as a function of $e_1$, which are derived from Eq.~\ref{eq:cuadratica} for each $m_1$ and $e_2$.  

As can be seen from Fig.~\ref{fig:multiplot_grafico1}, the GR plays a key role in defining the parameter space leading to oscillations of the orbital plane of an HZ test particle from prograde to retrograde inclinations and back again. In the absence of GR, orbital flips are associated exclusively with the nodal libration region, which depends only on $e_1$ \citep{Ziglin1975}. When GR effects are considered, orbital flips can be correlated with nodal librations as well as with nodal circulations \citep{Zanardi2018}. In this case, the study becomes more complex, and we distinguish three particular scenarios for discussion where the analysis of flipping particles undergoing nodal circulations is of special interest. On the one hand, GR can lead to scenarios in which nodal libration trajectories are not possible for any value of $e_1$, which is observed in the panels associated with $m_1$ = 1 M$_{\text{Sat}}$ and $e_2$ = 0.1 and 0.4 in Fig.~\ref{fig:multiplot_grafico1}. On the other hand, GR can generate a nodal libration region associated exclusively with purely retrograde orbits, which is shown in the panels associated with $m_1$ = 1 M$_{\text{Sat}}$ and $e_2$ = 0.7, and with $m_1$ = 1 M$_{\text{Jup}}$ and $e_2$ = 0.1 and 0.4 in Fig.~\ref{fig:multiplot_grafico1}. In all these cases, the study of flipping orbits with nodal circulations is of superlative interest, since they represent the only ones that lead to oscillations of the outer test particle’s orbital plane from prograde to retrograde inclinations and back again. Finally, GR can lead to nodal libration trajectories with orbital flips for particular values of $e_1$, which is observed in the panels associated with $m_1$ = 1 M$_{\text{Jup}}$ and $e_2$ = 0.7, and with $m_1$ = 3 M$_{\text{Jup}}$ and $e_2$ = 0.1, 0.4, and 0.7 in Fig.~\ref{fig:multiplot_grafico1}. In these cases, the study of flipping orbits with nodal circulations shows special interest since it extends the range of orbital parameters that lead to oscillations of the orbital plane of the outer test particle from prograde to retrograde inclinations and back again with respect to that obtained only by analyzing the nodal libration region in a given scenario.

Our results show that the more massive the inner perturber and the greater the value of $e_2$, the smaller the minimum $i^{\text{p}}_{2,\text{min}}$ that leads to nodal circulations with orbital flips for each $e_1$. Beyond this, we observe that the range of inclinations covered by nodal circulation trajectories with orbital flips shows different correlations with the parameters ($m_1$, $e_1$, $e_2$) depending on whether nodal libration trajectories with orbital flips are possible or not.

\begin{figure*}
   \centering
  \includegraphics[width=1
   \textwidth]{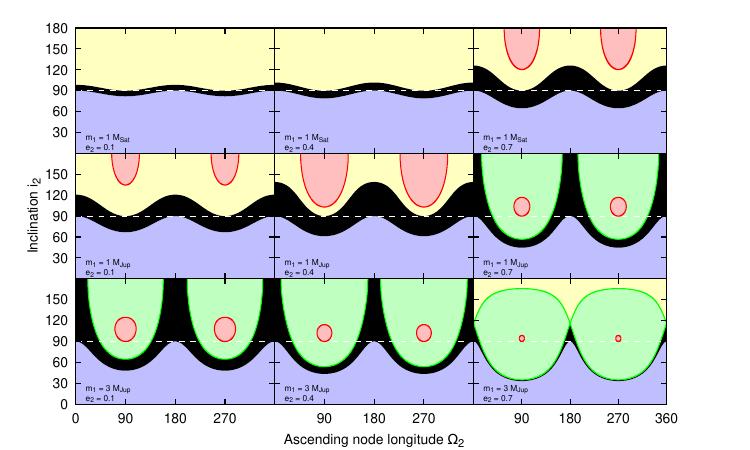}
   \caption{
Regimes of motion of an HZ test particle in the plane $\Omega_2$ vs. $i_2$ for $e_1$ = 0.7 and different values of $m_1$ and $e_2$. The references associated with the different regions and curves illustrated in each panel are described in
the caption of Fig.~\ref{fig:multiplot2_1}. 
}
\label{fig:multiplot2_3}
\end{figure*}

To understand this behavior, we present Figs~\ref{fig:multiplot2_1}, \ref{fig:multiplot2_2}, and \ref{fig:multiplot2_3}, which illustrate the different regimes of motion of an HZ test particle in a plane ($\Omega_2$, $i_2$) for $e_1$ = 0.1, 0.4, and 0.7, respectively, and different values of $m_1$ and $e_2$. From top to bottom, the rows of each of these figures are associated with a perturber of 1 M$_{\text{Sat}}$, 1 M$_{\text{Jup}}$, and 3 M$_{\text{Jup}}$, while the columns correspond to values of $e_2$ of 0.1, 0.4, and 0.7 from left to right. In each panel, on the one hand, the blue and yellow regions illustrate the pairs ($\Omega_2$, $i_2$) that lead to nodal circulations on prograde and retrograde orbits, respectively. On the other hand, the red and green regions represent the pairs ($\Omega_2$, $i_2$) that lead to nodal librations associated with purely retrograde orbits and orbital flips, respectively. The transition curve that divides the nodal libration trajectories with orbital flips of those associated with nodal circulations is illustrated in green. Moreover, the red curve delimits the nodal libration region with purely retrograde inclinations. Finally, the pairs ($\Omega_2$, $i_2$) leading to flipping orbits with nodal circulations are shown in black. 

From Fig.~\ref{fig:multiplot2_1}, nodal circulation trajectories with orbital flips for small $e_1$ show $i_2$ close to 90$^{\circ}$ for any $\Omega_2$, whatever the values of $m_1$ and $e_2$. According to Figs.~\ref{fig:multiplot2_2} and \ref{fig:multiplot2_3}, we find different results and correlations for moderate and high values of $e_1$, depending on whether the parameter space ($m_1$, $e_1$, $e_2$) allows or does not allow nodal libration trajectories with orbital flips.

For parameters ($m_1$, $e_1$, $e_2$) that do not lead to nodal libration trajectories with orbital flips, the region of the plane ($\Omega_2$, $i_2$) associated with nodal circulations with orbital flips increases with increasing $m_1$, $e_1$, and $e_2$. For parameters ($m_1$, $e_1$, $e_2$) that allow nodal libration trajectories with orbital flips, the region of the plane ($\Omega_2$, $i_2$) associated with nodal circulations with orbital flips increases with a decrease in $m_1$ and $e_2$, and with an increase in $e_1$.

It is important to remark that the results described have been derived from a secular theory up to the quadrupole level of the approximation with GR. In our scenarios of sudy, we assume a fixed semimajor axis ratio $a_1/a_2$ = 0.1 and values of $e_2$ = 0.1, 0.4, and 0.7. From \citet{Naoz2017}, an increase in $e_2$ can lead to a significant contribution from the octopole term in the secular Hamiltonian, which could cause the test particle's eccentricity to begin to oscillate and even grow beyond its initial value, promoting instabilities. From this, we decide to carry out N-body experiments with the aim of analyzing the validity of the equations provided in Sect.~\ref{sec:analytical_model} in our scenario of work. 

\subsection{Comparison with N-body experiments}
\label{subsec:N-body_experiments}

To test the robustness of our analytical prescriptions concerning the production of nodal circulation trajectories with orbital flips, we develop numerical experiments making use of IAS15, a 15th-order integrator with adaptive step-size control in the REBOUND N-body code \citep{Rein2015}.  In particular, GR effects are modeled using the $\text{gr}$ package included in REBOUNDx, which incorporates first-order post-newtonian effects by assuming that the masses are dominated by the central body of the system \citep{Tamayo2020}. 

All our numerical experiments assume a solar-mass star. The astrocentric orbit of the inner perturber of mass $m_1$ is characterized by the semimajor axis $a_1$, the eccentricity $e_1$, the inclination $i_1$, the argument of pericentre $\omega_1$, the ascending node longitude $\Omega_1$, and the true anomaly $\nu_1$. On the one hand, we adopt values of $m_1$ of 1 M$_{\oplus}$, 1 M$_{\textrm{Nep}}$, 1 M$_{\textrm{Sat}}$, 1 M$_{\textrm{Jup}}$, and 3 M$_{\textrm{Jup}}$. On the other hand, all numerical experiments assume initial values of $a_1$ = 0.1 au, $i_1$ = $\omega_1$ = $\Omega_1$ = 0$^{\circ}$, and $\nu_1$ = 180$^{\circ}$. Finally, $e_1$ adopt initial values of 0.1, 0.4, and 0.7, from which we developed Figs.~\ref{fig:multiplot2_1}, \ref{fig:multiplot2_2}, and \ref{fig:multiplot2_3}, respectively. The orbital parameters that define the barycentric orbit of the outer test particle are the semimajor axis $a_2$, the eccentricity $e_2$, the inclination $i_2$, the argument of pericentre $\omega_2$, the ascending node longitude $\Omega_2$, and the true anomaly $\nu_2$. For all cases, we adopt initial values of $a_2$ = 1 au, $\omega_2$ = 0$^{\circ}$, $\Omega_2$ = 90$^{\circ}$, and $\nu_2$ = 0$^{\circ}$. Moreover, our numerical experiments are run by assuming initial $e_2$ of 0.1, 0.4, and 0.7, which is in agreement with the values adopted in the different panels of Figs.~\ref{fig:multiplot2_1}, \ref{fig:multiplot2_2}, and \ref{fig:multiplot2_3}. Finally, for each particular scenario, the initial $i_2$ at $\Omega_2$ = 90$^{\circ}$ are selected from the values of $i^{\textrm{p}}_{2,\textrm{min}}$ that are solutions of Eq.~\ref{eq:cuadratica_flipping}, which lead to nodal circulations with orbital flips of an outer test particle according to a secular theory up to the quadrupole level with GR. To carry out a comparative analysis with the results derived from analytical theory described in Sect.~\ref{sec:analytical_model}, the orbital elements of the outer test particle obtained from the N-body experiments must be referenced to barycentre and invariant plane of the system, where x-axis is directed toward the pericentre of the inner perturber. Due to the GR effects, the orbit of the inner perturber experiences apsidal precession, for which $\Omega_2$ is measured with respect to a rotating system. Finally, we integrate all N-body simulations for a total time of 1 Myr, which corresponds to about 3.1 $\times$ 10$^{7}$ orbital periods of the inner perturber and 9.46, 11.15, and 18.36 periods of apsidal precession of the inner orbit due to GR for values of $e_1$ = 0.1, 0.4, and 0.7, respectively.

For an inner Earth- and Neptune-mass planet, we find a very good agreement between the evolution of the outer test particles derived from the N-body experiments and the analytical theory. In fact, each test particle experiences nodal circulations with orbital flips with values of $i_2$ close to 90$^{\circ}$ for any $\Omega_2$, while $e_2$ evolves with very low-amplitude oscillations around its initial value, which is consistent with the results of the secular theory up to the quadrupole level. 

For an inner Saturn-mass planet, the results derived from the N-body simulations agree very well with the analytical approach for $e_1$ = 0.1, 0.4, and 0.7, and initial $e_2$ = 0.1 and 0.4. For initial $e_2$ = 0.7, the evolutionary trajectory of a test particle obtained from a N-body simulation is consistent with that derived from analytical theory, although some deviations are observed, which become more evident with increasing $e_1$. In particular, for $e_1$ = 0.7, two particles with an initial $e_2$ = 0.7 are ejected from the system in our N-body experiments due to significant increases in their orbital eccentricities. 

\begin{figure}
   \centering
  \includegraphics[width=0.5
   \textwidth]{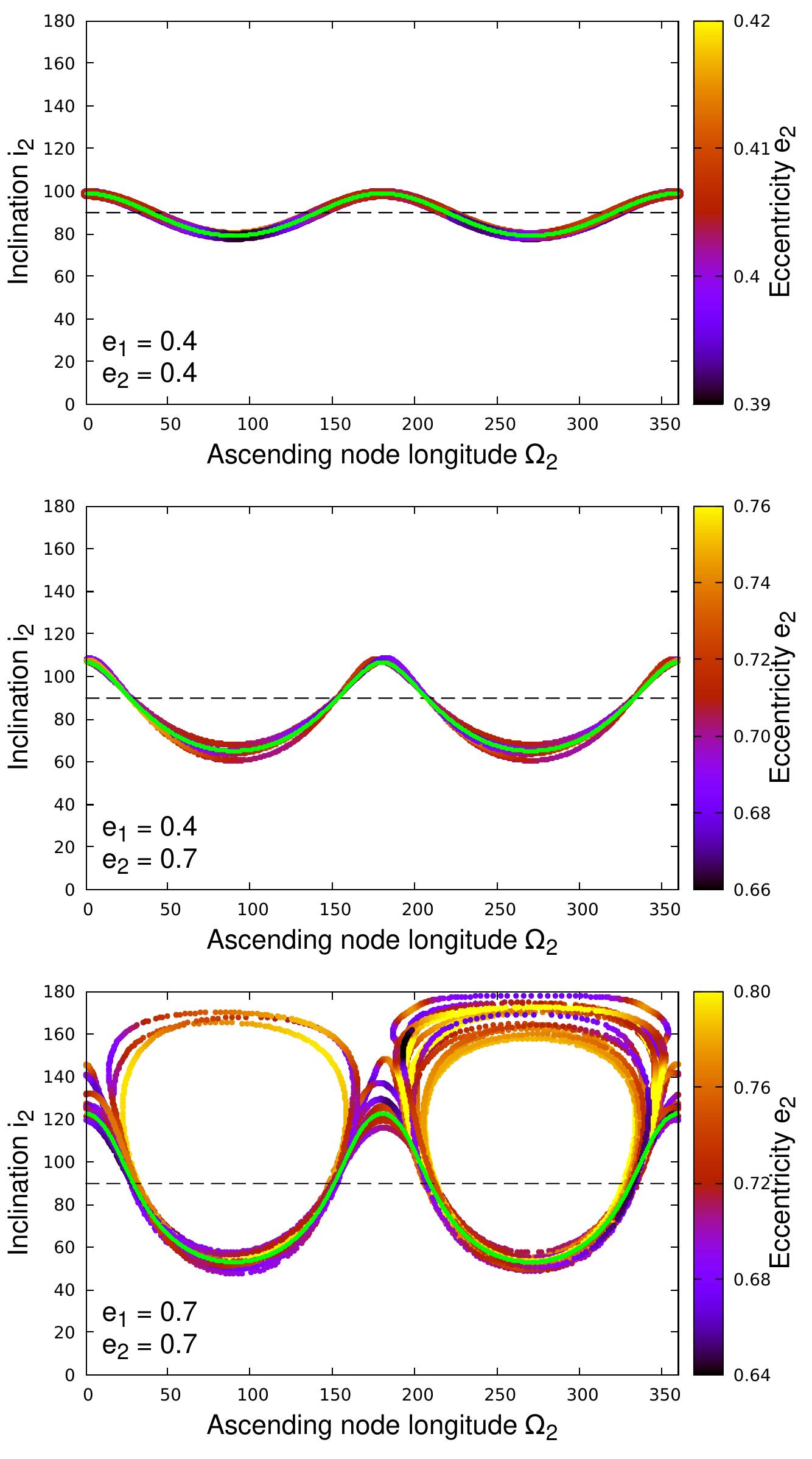}
   \caption{Evolutionary trayectories of HZ test particles in the plane $\Omega_2$ vs. $i_2$ obtained from N-body experiments (color points) and the analytical theory (green points), where the color palette illustrates the change in the orbital eccentricity $e_2$ throughout 1 Myr. In all cases, we assume $m_\star$ = 1 M$_\odot$, $m_1$ = 1 M$_{\textrm{Jup}}$, and initial orbital parameters of $a_1$ = 0.1 au, $i_1$ = $\Omega_1$ = $\omega_1$ = 0$^{\circ}$, $\nu_1$ = 180$^{\circ}$, $a_2$ = 1 au, $\Omega_2$ = 90$^{\circ}$, and $\omega_2$ = $\nu_2$ = 0$^{\circ}$. In particular, we assume initial values of $e_1$ = 0.4, $e_2$ = 0.4, $i_2$ = 79$^{\circ}$ in the top panel, $e_1$ = 0.4, $e_2$ = 0.7, $i_2$ = 65$^{\circ}$ in the middle panel, and $e_1$ = 0.7, $e_2$ = 0.7, $i_2$ = 53$^{\circ}$ in the bottom panel.
}
\label{fig:multiplot3}
\end{figure}

For an inner Jupiter-mass planet, we observe that the trajectories of nodal circulations with orbital flips of outer test particles with initial values of $e_2$ = 0.1 and 0.4 resulting from N-body experiments are in a very good agreement with the analytical theory for $e_1$ = 0.1, 0.4, and 0.7. The top panel of Fig.~\ref{fig:multiplot3} illustrates an excellent consistency between the trajectories of an outer test particle in the plane ($\Omega_2$, $i_2$), which result from an N-body simulation (color points) and the analytical theory (green points) for an initial $e_2$ = $e_1$ = 0.4. Moreover, the color code shows that $e_2$ does not experience significant changes over 1 Myr. For the particular case of $e_1$ = 0.7 and an initial $e_2$ = 0.4, the evolution of a test particle derived from an N-body simulation shows some deviations from that obtained from the analytical theory for $i_2$ at $\Omega_2$ = 90$^{\circ}$ higher than about 85$^{\circ}$. In this particular scenario, the particles simulated in our N-body experiments are ejected from the system for $i_2$ at $\Omega_2$ = 90$^{\circ}$ close to 90$^{\circ}$. 

For an inner Jupiter-mass planet, the evolution of the test particle is more complex for an initial $e_2$ = 0.7. For $e_1$ = 0.1, the trajectories of nodal circulation with orbital flips of outer test particles resulting from N-body simulations are in a good agreement with the analytical theory for initial $i_2$ at $\Omega_2$ = 90$^{\circ}$ less than about 89$^{\circ}$. The most significant discrepancies are observed for an initial $i_2$ at $\Omega_2$ = 90$^{\circ}$ between about 89$^{\circ}$ and 90$^{\circ}$, where the evolutionary trajectory resulting from the N-body simulation shows that the nodal circulations of the test particles are associated with both orbital flips and purely retrograde orbits over 1 Myr. Beyond this small deviations, $i_2$ always adopts values close to 90$^{\circ}$ for any $\Omega_2$ throughout its evolution, which is consistent with analytical theory. For $e_1$ = 0.4 and 0.7, the region of nodal circulation with orbital flips is delimited by the nodal libration region for $e_2$ = 0.7. This can be observed in the right middle panel of Figs.~\ref{fig:multiplot2_2} and \ref{fig:multiplot2_3}, where the black and green zones illustrate the pairs ($\Omega_2$, $i_2$) that lead to orbital flips correlated with nodal circulations and nodal librations, respectively, and the green curve represents a transition between both regimes. For $e_1$ = 0.4 and $e_2$ = 0.7, the solutions of Eq.~\ref{eq:cuadratica_flipping} shows that the $i_2$ at $\Omega_2$ = 90$^{\circ}$ that define the nodal circulation region with orbital flips range from 60.47$^{\circ}$ to 65.94$^{\circ}$, where this last value is associated with the curve that transitions from nodal circulations to nodal librations. From an analysis of such scenarios, we observed that the analytical and N-body simulation results are consistent for initial $i_2$ at $\Omega_2$ = 90$^{\circ}$ less than about 65.5$^{\circ}$, although the evolutionary trajectories of the test particles in the plane ($\Omega_2$, $i_2$) derived from numerical experiments show deviations with respect to those derived from the analytical theory. A comparison between these trajectories can be seen in the middle panel of Fig.~\ref{fig:multiplot3}. As the reader can see from the top and the middle panels of such figure, the differences between the N-body experiments and the analytical theory are more noticeable with increasing $e_2$. For initial $i_2$ at $\Omega_2$ = 90$^{\circ}$ selected between about 65.5$^{\circ}$ and the curve that divides nodal circulations and nodal librations, the evolution of the test particles resulting from N-body simulations are not in agreement with the analytical theory since the temporal evolution of $\Omega_2$ frequently switches between libration and circulation throughout 1 Myr. For $e_1$ = $e_2$ = 0.7, Eq.~\ref{eq:cuadratica_flipping} shows that the nodal circulation region with orbital flips is defined by $i_2$ at $\Omega_2$ = 90$^{\circ}$ ranging from 44.68$^{\circ}$ to 56.76$^{\circ}$, where this last value refers to the curve that transitions from nodal circulations to nodal librations. From this, the N-body simulation results are in a good agreement with our analytical criteria for an initial $i_2$ at $\Omega_2$ = 90$^{\circ}$ less than about 50.25$^{\circ}$, although the trajectories resulting from the N-body experiments show some deviations respect to that derived from analytical theory, which is similar to what was described for the scenario associated with $e_1$ = 0.4 and $e_2$ = 0.7. For initial values of $i_2$ at $\Omega_2$ = 90$^{\circ}$ between about 50.25$^{\circ}$ and the curve that divides nodal circulations and nodal librations, the consistency between the analytical and N-body simulation results is not good since the test particles experience circulations and librations of $\Omega_2$ over 1 Myr, with evident changes in the orbital eccentricity. An example of such a behavior is illustrated in the bottom panel of Fig.~\ref{fig:multiplot3}. For $e_2$ = 0.7, a comparison between the scenarios assuming values of $e_1$ of 0.4 and 0.7 shows that an increase in $e_1$ leads to a decrease in the range of values of $i_2$ that are solutions of Eq.~\ref{eq:cuadratica_flipping} and produce a good agreement between the evolution of a test particle resulting from an N-body experiment and that derived from the analytical theory.

For an inner super-Jupiter with $m_1$ = 3 M$_{\text{Jup}}$, the comparison between the N-body experiments and the analytical theory shows some similarities and differences with respect to that described for an inner Jupiter-mass planet. For $e_1$ = 0.1, the evolution of the test particles resulting from N-body experiments are in a very good agreement with the analytical theory for initial $e_2$ of 0.1 and 0.4. For $e_2$ = 0.7, the analytical model shows that the nodal circulation region with orbital flips is delimited by the nodal libration region, which can be observed in the right bottom panel of Fig.~\ref{fig:multiplot2_1}. From Eq.~\ref{eq:cuadratica_flipping}, the nodal circulation region with orbital flips is very narrow and it is defined by $i_2$ at $\Omega_2$ = 90$^{\circ}$ ranging from 82.3$^{\circ}$ to 83.87$^{\circ}$, where this last value corresponds to the transition curve between nodal circulations and nodal librations. By selecting initial $i_2$ from such a range of values, we find that the N-body simulations are not in agreement with the analytical theory. In fact, the test particles of such scenario experience librations and circulations of $\Omega_2$ throughout 1 Myr or escape from the system due to significant excitations of the orbital eccentricity.

For $e_1$ = 0.4 and 0.7, the nodal circulation region with orbital flips is delimited by the nodal libration region for $e_2$ = 0.1, 0.4, and 0.7, such as is illustrated in the bottom panels of Figs.~\ref{fig:multiplot2_2} and \ref{fig:multiplot2_3}. On the one hand, for an initial $e_2$ = 0.1, the results derived from N-body simulations are in a very good agreement with the analytical theory for initial values of $i_2$ at $\Omega_2$ = 90$^{\circ}$ selected from the full range of solutions of Eq.~\ref{eq:cuadratica_flipping}, even those that are comparable to the value of $i_2$ at $\Omega_2$ = 90$^{\circ}$ associated with the transition curve between nodal circulations and nodal librations. On the other hand, for an initial $e_2$ = 0.4, the evolution of test particles resulting from the N-body experiments is consistent with that derived from analytical theory, although some small deviations are evident. Such a consistency is observed for initial values of $i_2$ at $\Omega_2$ = 90$^{\circ}$ corresponding to the set of solutions of Eq.~\ref{eq:cuadratica_flipping} that are smaller than a given limiting value. For $e_1$ = 0.4, such a limiting value of $i_2$ at $\Omega_2$ = 90$^{\circ}$ is about 63.85$^{\circ}$, which is very close to the value associated with the transition curve between nodal circulations and nodal librations, which is of 64.04$^{\circ}$. For $e_1$ = 0.7,  such a limiting value of $i_2$ at $\Omega_2$ = 90$^{\circ}$ is about 49.95$^{\circ}$, which is smaller than the value associated with the transition curve between nodal circulations and nodal librations, which is of 53.91$^{\circ}$. In all these cases, the results derived from the N-body simulations do not agree with the analytical theory for $i_2$ at $\Omega_2$ = 90$^{\circ}$ ranging between the specified limit value and that associated with the transition curve between nodal circulations and nodal librations. For this space of parameters, the numerical experiments show that the evolution of $\Omega_2$ frequently switches between libration and circulation throughout 1 Myr. For $e_2$ = 0.4, we observe that an increase in $e_1$ leads to a decrease in the range of values of $i_2$ that are solutions of Eq.~\ref{eq:cuadratica_flipping} and produce good agreements between the analytical theory and the N-body experiments. Finally, for an initial $e_2$ = 0.7, the N-body experiments are not in agreement with the analytical theory for the entire range of solutions of Eq.~\ref{eq:cuadratica_flipping}. On the one hand, for $e_1$ = 0.4, the evolution of the test particles derived from the numerical simulations shows that $\Omega_2$ switches between libration and circulation throughout 1 Myr. On the other hand, for $e_1$ = 0.7, all simulated particles escape from the system due to significant increases in eccentricity. 

\section{Discussion and Conclusions}
\label{sec:conclusions}

We analyzed the role of the GR in the dynamics of an outer test particle that experiences nodal circulations with orbital flips in the elliptical restricted three-body problem. In particular, the central object of our systems of work is a solar-mass star and the test particle is assumed to be located at the HZ with a semimajor axis of $a_2$ = 1 au, which evolves under the effects of an inner perturber of planetary mass with a semimajor axis of $a_1$ = 0.1 au. 

Working on the basis of the secular theory up to the quadrupole level with GR developed by \citet{Zanardi2018} and \citet{Zanardi2023}, we derived analytical expressions that allow to calculate the inclinations that lead to an outer test particle to evolve on trajectories of nodal circulation with orbital flips. In particular, we analyze the sensitivity of our results to the mass $m_1$ and eccentricity $e_1$ of the inner perturber, as well as to the eccentricity $e_2$ of the outer test particle. 

Our results show that nodal circulation trajectories with orbital flips of an HZ test particle with eccentricity $e_2$ are possible for any mass $m_1$ and eccentricity $e_1$ of the inner perturber and a suitable inclination $i_2$. In particular, for $m_1 \lesssim$ 1 M$_{\textrm{Nep}}$, nodal circulation trajectories with orbital flips of HZ test particles have $i_2$ close to 90$^{\circ}$ for all $\Omega_2$ between 0$^{\circ}$ and 360$^{\circ}$, and for any value of $e_1$ and $e_2$. For $m_1 \gtrsim$ 1 M$_{\textrm{Sat}}$ and small $e_1$, HZ test particles experience nodal circulations with orbital flips, where $i_2$ keep values close to 90 for all $\Omega_2$ and for any $e_2$. In this line of research, we find that the greater the values of $m_1$ and $e_2$, the smaller the minimum $i^{\textrm{p}}_{2,\textrm{min}}$ capable of producing nodal circulations with orbital flips for each $e_1$. Moreover, if the set of parameters ($m_1$, $e_1$, $e_2$) does not allow nodal libration trajectories with orbital flips for any $i_2$, the region of the plane ($\Omega_2$, $i_2$) associated with nodal circulations with orbital flips increases with increasing $m_1$, $e_1$, and $e_2$. If nodal librations with orbital flips occur, the region of the plane ($\Omega_2$, $i_2$) referred to nodal circulations with orbital flips increases with a decrease in $m_1$ and $e_2$, and with an increase in $e_1$.

Although this work has focused on analyzing nodal circulation trajectories with orbital flips of massless HZ particles evolving under the influence of an inner perturber of planetary mass around a solar-mass star, considering GR effects, the analytical equations presented can be used to analyze the secular dynamics at the quadrupole level of an outer massless particle around any binary system of 
massive bodies.

We carry out N-body simulations aimed at analyzing the robustness of the analytical criteria that lead to nodal circulations with orbital flips of HZ test particles due to GR effects. In general terms, our analysis shows that the N-body experiments are consistent with the analytical theory for $m_1$ ranging from Earth-mass planets to super-Jupiters and for $e_2$ = 0.1 and 0.4, assuming from low, to moderate, and even high values of $e_1$. For $m_1 \lesssim$ 1 M$_\textrm{Jup}$, such a consistency is also observed for $e_2$ = 0.7, although an increase in $m_1$ and $e_1$ leads to a decrease in the range of $i_2$ that produces nodal circulations with orbital flips from the analytical theory and evidences good agreements with the N-body simulations. Finally, for $m_1$ = 3 M$_{\textrm{Jup}}$, the N-body experiments do not agree with the analytical theory for $e_2$ = 0.7, and $e_1$ = 0.1, 0.4, and 0.7.

It is important to remark that the present research is aimed to analyzing the dynamics of outer massless particles in the elliptical restricted three-body problem with GR effects, which produce the precession of the argument of pericentre of the inner orbit associated with the massive bodies of the system. We believe that a discussion of the assumptions made in our model could lead the reader to a better interpretation of the results, as well as to a correct reasoning concerning its range of application. From \citet{Zanardi2018}, the GR produces fixed points associated with retrograde orbital inclinations of an outer massless particle in the elliptical restricted three-body problem. In addition to GR, the apsidal precession of the inner orbit could also arise from other effects such as tides and rotation-induced flattening \citep{Sterne1939}, which would lead to a modification of the inclination associated with the pixed points of the system and from this to new limits corresponding to the different regimes of motion of the test particle.  Moreover, \citet{Lidov1976}, \citet{Ferrer1994}, \citet{FaragoLaskar2010}, and \citet{Zanazzi2018} analyzed the problem of three massive bodies with classical Newtonian gravitation and showed that if the outer object were sufficiently massive, the fixed points would be associated  with prograde inclinations. Due to all these considerations, it is important to remark that the combined action of the aforementioned effects could lead to significant changes in the analytical prescriptions derived in the present investigation and, consequently, modify the criteria that define the nodal circulation trajectories with orbital flips of an outer particle in a three-body problem. Although the study of the influence of such effects is beyond the scope of this work, we will address such research in future articles.

The present investigation, together with the work developed by \citet{Coronel2024}, offers a complete description of the role of GR in the secular dynamics at the quadrupole level of HZ massless particles, which evolve under the influence of a planetary mass perturber with a semimajor axis of 0.1 au orbiting around a solar-mass star. Studies of this kind aim to deepen our understanding of the dynamics of such a distinctive region of a planetary system as the habitable zone.

\begin{acknowledgements}
We thank the anonymous referee for her/his comments and suggestions, which have helped us improve the manuscript. This work was partially financed by Universidad Nacional de La Plata, Argentina, through the PID G172. Moreover, the authors acknowledge the partial financial support by Facultad de Ciencias Astronómicas y Geofísicas de la Universidad Nacional de La Plata, and Instituto de Astrofísica de La Plata, for extensive use of their computing facilities.
\end{acknowledgements}

\bibliographystyle{raa}
\bibliography{Paper}

\label{lastpage}

\end{document}